\documentstyle[aps,prb,multicol]{revtex}

\newcommand{\beq}{\begin{eqnarray}}
\newcommand{\eeq}{\end{eqnarray}}
\newcommand{\n}{\nonumber}
\newcommand{\epi}{\epsilon_\infty}
\begin{document}
\draft
\title
{Johnson-Nyquist noise in narrow wires}
\author{ Misha Turlakov}
\address{ Department of Physics, University of Illinois, 1110 W. Green Street, Urbana, IL 61801}
\date{\today}
\maketitle

\begin{abstract}

The Johnson-Nyquist noise in narrow semiconducting wires having a transverse size smaller than
the screening length is shown to be white up to frequency $D/L^2$ and to decay
at higher frequencies as $\omega^{-\frac{1}{2}}$. This result is contrasted
with the noise spectra in neutral and charged liquids.

\end{abstract}

\begin{multicols}{2}


 It is interesting to compare properties of charged and neutral systems.
The role of the Coulomb interaction 
crucially depends on the effective dimensionality
of the charged system. For instance, due to the long-range nature of the
Coulomb interaction in three-dimensional systems the density excitations of neutral liquids
(acoustic phonons) are transformed to gapped plasmons, while in one- and two-dimensional
systems plasmons remain gapless.   
Here I examine the noise spectrum as  another aspect 
of the singular role of Coulomb interaction critically depending on the dimensionality.

The noise spectrum is quite different in charged and neutral liquids. The equilibrium Johnson-Nyquist
noise\cite{Nyquist}
 in an electrical conductor (with a screening length smaller than any size of the conductor)
 is white up to very high frequencies
(the smaller of the elastic scattering rate $1/\tau$ and the Maxwell relaxation frequency $4\pi\sigma$);
while in neutral liquids, the noise becomes frequency-dependent above the ``Thouless'' frequency
$D/L^2$ ($D$ is a diffusion constant and $L$ is the distance between points). The difference 
is due to screening in charged liquids and depends on the dimensionality of
the conductor. I show here that
for electrical wires having a transverse size $a$ smaller than the Debye screening 
length $\lambda_D$ (noted here as narrow wires), the Johnson-Nyquist noise decays 
as  $\omega^{-\frac{1}{2}}$ above the ``Thouless''
frequency {\it independently of external screening}.
 The threshold frequency $D/L^2$ can be made arbitrarily small by increasing the separation
between two points $L$.


To calculate the fluctuations of the electrochemical potential,
 we need to relate it to  the coupled
fluctuations of charge density and currents. We start by writing the continuity equation and
the current equation valid in the hydrodynamic limit\cite{Nozieres}:

\beq
\frac{\partial \rho}{\partial t} + div(\vec{j})=0;~~\vec{j}=
\sigma \vec{E}^{tot} - D \vec{\nabla} \rho. \n
\eeq

For self-consistency, we need to account for the potential induced by the fluctuation of charge density:
$\phi_{q,\omega}^{ind}= \frac{u_1(q)}{\epi}  \rho_{q,\omega}$ (Coulomb's law). 
Since we consider a conductor
with transverse dimensions $a$ smaller than the screening length $\lambda_D$, $u_1(q)= 2 ln\frac{1}{qa}$ is
a one-dimensional Coulomb potential ($q$ is a wave vector along a one-dimensional conductor).
The total potential driving current
is the sum of the external and induced potentials. Thus the full system of equations is: 

\beq
i\omega \rho_{q,\omega}+iq j_{q,\omega}=0,~~j_{q,\omega}=\sigma(iq) \phi_{q,\omega}^{tot}
+D(iq)\rho_{q,\omega}, \n \\
\phi_{q,\omega}^{tot}=\phi_{q,\omega}^{ext}+\phi_{q,\omega}^{ind},
~~\phi_{q,\omega}^{ind}= \frac{2 ln\frac{1}{qa}}{\epi} \rho_{q,\omega}. \n
\eeq

Finally, after some elementary algebra, we can use the above equations to relate the charge density 
variation to the external potential:

\beq
\rho_{q,\omega}=-\frac{\sigma_1 q^2}{-i\omega+
(D+2 (\sigma_1/\epi)ln(1/qa))q^2}\phi_{q,\omega}^{ext}. \n 
\eeq 

Using the Einstein relation $\sigma=D\chi_0$ and 
the expression for the one-dimensional conductivity $\sigma_1=\sigma a^2$, 
the density-density response function is

\beq
\chi_{q,\omega} \equiv - \frac{D\chi_0 a^2 q^2}{-i\omega+Dq^2(1 +(2a^2\chi_0/\epi)ln(1/qa)) }.  \n
\eeq

We can now apply the fluctuation dissipation theorem (FDT) to calculate the density fluctuation
spectrum (assuming classical fluctuations,  $\hbar \omega \ll  kT$):

\beq
<\mid\delta \rho_{q,\omega} \mid^2>=\hbar Im\chi_{q,\omega} coth(\frac{\hbar \omega}{2kT}) \cong
\frac{2 kT}{\omega} Im\chi_{q,\omega}.  \n
\eeq

The static charge compressibility $\chi_0$ is simply related to the Debye screening (or Thomas-Fermi)
length: $1/\lambda_D^2=4\pi\chi_0$. 
The induced potential fluctuations can be expressed through the charge density fluctuations:

\beq
<\mid \phi_{q,\omega}^{ind} \mid^2>= \frac{(u_1(q))^2}{\epi^2} <\mid\delta \rho_{q,\omega} \mid^2>, \n \\
<\mid \phi_{q,\omega}^{ind} \mid^2>= 
\frac{ 2 kT }{\epi^2 }
\frac{\sigma_1 q^2(2ln\frac{1}{qa})^2}
{\omega^2 +D^2(1+\frac{a^2}{2\pi\epi\lambda_D^2}ln\frac{1}{qa})^2q^4}.
\label{eq:1d} 
\eeq


We can compare Eqn. (\ref{eq:1d}) with the spectral density of potential fluctuations
in bulk three-dimensional charged and neutral liquids.
In the case of a three-dimensional charged liquid, we need to
use the three-dimensional Coulomb potential 
$\phi_{q,\omega}^{ind}=\frac{u_3(q)}{\epi}\rho_{q,\omega} =\frac{4 \pi}{\epi q^2} \rho_{q,\omega}$.
Following the above simple derivation, we get the expression for voltage fluctuations
(it is sufficient for our purposes to consider only longitudinal fluctuations) in a three-dimensional
conductor\cite{2-der}:

\beq
<\mid \phi_{q,\omega}^{ind(3d)} \mid^2>= 
\frac{32 \pi^2 kT}{\epi^2 q^2}\frac{\sigma}{\omega^2 +(Dq^2+4 \pi\sigma/\epi)^2}. \label{eq:charged}
\eeq

 In the case of a neutral liquid, there is no
long-range induced potential; therefore, we get the standard density-density response function
and potential fluctuations describing diffusion:

\beq
<\mid \phi_{q,\omega}^{(n)} \mid^2>= \frac{<\mid\delta \rho_{q,\omega} \mid^2>}{\chi_0^2}=
\frac{2T}{\chi_0^2} \frac{D |\chi_0| q^2}{\omega^2 +(Dq^2)^2}. \label{eq:neutral}
\eeq

We can now use the spectral densities (Eqns. \ref{eq:1d}-\ref{eq:neutral}) to calculate the
experimentally measured differential noise between the two ends of the sample, averaged
over transverse modes:

\beq
<\mid \phi_{12}(\omega) \mid^2>= 
\sum_{q_x} \frac{ sin^2(q_x a)}{q_x^2 a^2}
\sum_{q_y} \frac{ sin^2(q_y a)}{q_y^2 a^2} \n \\
\int dq_z~ 4sin^2 (\frac{q_zL}{2}) 
<\mid \phi (q,\omega) \mid^2>. \label{eq:ends} 
\eeq

The Johnson-Nyquist noise in a three-dimensional conductor (\ref{eq:charged}) can be easily calculated,
because the dominant contribution to the sums and the integral comes from ``zero modes''
($q_x,q_y,q_z \rightarrow 0$):

\beq
<\mid \phi_{12}^{3d}(\omega) \mid^2> = 2kT \frac{16\pi^2\sigma L}{\epi^2 S} 
\frac{1}{(\frac{4\pi\sigma}{\epi})^2 + \omega^2}. 
\eeq

Such noise is readily interpreted as the noise from a conductor 
 having an internal resistance $R=\frac{L}{\sigma S}$ and an
internal capacitance $C=\frac{\epsilon S}{4\pi L}$ connected in parallel\cite{Rytov}:
$ R(\omega)=R/(1+(RC\omega)^2)$. Remarkably, the Johnson-Nyquist noise is white up to the frequency
$4\pi\sigma$, which is independent of the length of the wire. The appropriate physical picture
of fluctuations in an electrical conductor is that charge fluctuations relax on a very fast
time scale $1/4\pi\sigma$, producing quasi-homogeneous current fluctuations. It is important to point out
that the noise can depend on frequency through the frequency dependence of the conductivity $\sigma(\omega)$.
For the Drude model of conductivity, the characteristic frequency for fall-off of the conductivity
$\sigma(\omega)$ is then the elastic scattering rate $1/\tau$. 

In the case of a ``one-dimensional'' wire $\lambda_D > a$, we can take into account only one
``zero mode''($q_x=q_y=0$), since higher harmonics make contributions smaller in powers of $(a/L)^2$.
If we approximate the weak logarithmic dependence in Eqn.(\ref{eq:1d}) by a constant 
$ln\frac{1}{qa} \rightarrow ln\frac{L}{a}$,
we get an expression similar to Eqn.(\ref{eq:neutral}) with the renormalized diffusion
coefficient $D'\equiv D(1+\frac{a^2}{2\pi\epsilon\lambda_D^2} ln(L/a))$.
 Thus the frequency dependence of noise
for a ``one-dimensional'' wire is the same as for a neutral liquid.
 This result is expected, since the screening is not efficient in one dimension.  
The integral over wave vector $q_z$ can be evaluated explicitly, while the approximation of
the logarithm by a constant is reliable for limits $\omega \rightarrow 0$ and $\omega \rightarrow \infty$.

\beq
<\mid \phi_{12}^{1d}(\omega) \mid^2> = 2kT \frac{D^2}{D'^2} \frac{\pi L}{\sigma_1 \theta}
(1-e^{-\theta}(cos\theta-sin\theta)), 
\eeq
where $\theta=(\omega/2\omega_0)^{1/2}$ and $\omega_0=D'/L^2$ is the natural diffusion frequency.
From the above expression for noise in a one-dimensional wire, we see that it is white
up to the ``Thouless'' frequency $\omega_0$, and it decays above this frequency as $1/\sqrt{\omega}$.
The same frequency dependence (with $D' \equiv D$) is expected 
for the fluctuations of the chemical potential
between two points in a  narrow vessel ($\omega \ll D/a^2$) of liquid.
 It is clear indeed that the difference
in chemical potentials between two points is relaxed through diffusion on a  characteristic time scale
$L^2/D$.  
In fact, it is the classical result for any quantity (such as temperature, density) obeying
a diffusion process that does not have long-range correlations.\cite{voss} 


The nature of the relaxation of a random potential fluctuation is quite different in charged
and neutral liquids. In charged liquids, it is essentially the fast process of screening, and
in neutral liquids it is the process of diffusion. In  reduced dimensionality systems (such as narrow wires),
the Coulomb interaction does not cause long-range correlations; therefore, the noise in a narrow conductor
can be similar to neutral systems.

The experimental observation of predicted noise properties is feasible in semiconducting materials
having a low concentration of carriers.\cite{validity} The screening length $\lambda_D$ in such materials 
can be as large as $10^{-4} cm$. In fact, with current experimental techniques 
(see Reference\cite{Naveh} for a review of experiments), even the high Maxwell relaxation frequency
crossover can be observed in ``wide'' wires ($a \gg \lambda_D$, the situation almost always encountered)
 with poor conductivity.
In metals, normally both the elastic rate $1/\tau$ and the Maxwell frequency $4\pi\sigma$ are
high and difficult to observe.

The question of the frequency dependence of equilibrium and ``shot'' noise  
was raised recently by Y. Naveh {\it et al}.\cite{Naveh} Special geometries (sandwich and ground plane)
were suggested to observe the Maxwell and Thouless crossover frequencies. The above calculation shows
that the crossover at the Maxwell relaxation frequency is a general property of Coulomb systems
 and should be observed independently of geometry and length $L$ for ``wide'' wires. 
Moreover, for ``narrow'' wires ($\lambda_D > a$)
the Thouless frequency crossover should be seen independently of ``external screening'' by
electrodes or the ground plane.

In conclusion, the noise in narrow conductors ($\lambda_D > a$) becomes frequency-dependent
starting from low frequency $D/L^2$ (quite similar to simple diffusion systems), although
in wide conductors, the noise is white up to the smaller of the frequencies $4\pi\sigma$ or $1/\tau$.

This work was supported by the National Science Foundation through the Science and Technology
Center for Superconductivity (Grant No. DMR-91-20000).
I thank  A. Leggett and M. Weissman for helpful discussions.


\end{multicols}


\begin{references}
\bibitem{2-der}
Equation (\ref{eq:charged}) can be also derived in two other ways, elucidating
the meaning of charge and current fluctuations in a charged liquid. 
One way is to introduce
a stochastic current source $j^{ac}$ 
into the current equation\cite{kogan}:
 
\beq
j_{q,\omega}=\sigma(iq) \phi_{q,\omega}
+D(iq)\rho_{q,\omega} + j^{ac} \n   
\eeq

This derivation illustrates the role of quasi-homogeneous current fluctuations effectively
screening charge fluctuations. Yet a third derivation explicitly demonstrates the role
of screening:

\beq
\delta \phi^{tot}_{q,\omega}= \frac{4\pi}{\epsilon(q,\omega)\epsilon_0 q^2} \delta \rho_{q,\omega} \n
\eeq 

The fluctuations of $\delta \phi^{tot}_{q,\omega}$ are
given by Eqn.(\ref{eq:charged}) with the dielectric constant  $\epsilon(q,\omega)$ calculated
in the hydrodynamic limit\cite{Nozieres}:

\beq
\epsilon(q,\omega)= 1-\frac{4\pi}{\epsilon_\infty q^2} \chi^{(n)}(q,\omega)=
1+\frac{4 \pi \sigma}{(i\omega +D q^2) \epsilon_\infty}. \n
\eeq
\bibitem{validity}
It is important to point out that conditions $\sigma=D/4\pi\lambda_D^2 \gg e^2/\hbar$ and
$q_F \gg \pi/a$ ($q_F \simeq (\pi/4)(\hbar^2/e^2m)(1/\lambda_D^2)$ can be satisfied while
 $a \ll \lambda_D$, so that ``weak localization'' and Luttinger-liquid  physics are not relevant
for the problem discussed here.
\end{references}
\end{document}